\documentclass[prb,twocolumn,aps,showpacs,10pt,longbibliography]{revtex4-1}

\usepackage[pdftex]{graphicx} % Include figure files      
\usepackage{epstopdf}
\usepackage{verbatim}
\usepackage{color}
\usepackage{subfigure}
\usepackage{tabularx}
\usepackage{amsfonts}
\usepackage{wasysym}
\usepackage{bm}
\usepackage{float}
\usepackage{bbold}

\usepackage{nicefrac}

\newcommand{\avg}[1]{\langle\,#1\,\rangle}

\newcolumntype{C}[1]{>{\centering\let\newline\\\arraybackslash\hspace{0pt}}m{#1}}

\begin{document}
\title{The effect of spin-orbit coupling on 
the effective-spin correlation in YbMgGaO$_4$}
\author{Yao-Dong Li$^{1}$}
\author{Yao Shen$^{1}$}
\author{Yuesheng Li$^{2,3}$}
\author{Jun Zhao$^{1,4}$}
\author{Gang Chen$^{1,4}$}
\email{gangchen.physics@gmail.com, gchen$_$physics@fudan.edu.cn}
\affiliation{$^{1}$State Key Laboratory of Surface Physics, 
Center for Field Theory and Particle Physics,
Department of Physics, Fudan University,
Shanghai 200433, People's Republic of China}
\affiliation{$^{2}$Department of Physics, Renmin University of China, 
Beijing 100872, People's Republic of China}
\affiliation{$^{3}$Experimental Physics VI, Center for Electronic Correlations and Magnetism, 
University of Augsburg, 86159 Augsburg, Germany}
\affiliation{$^{4}$Collaborative Innovation Center of Advanced Microstructures,
Nanjing, 210093, People's Republic of China}

\date{\today}

\begin{abstract}
Motivated by the recent experiments on the triangular lattice spin liquid 
YbMgGaO$_4$, we explore the effect of spin-orbit coupling on the effective-spin 
correlation of the Yb local moments. We point out the anisotropic interaction between 
the effective-spins on the nearest neighbor bonds is sufficient to reproduce the 
spin-wave dispersion of the fully polarized state in the presence of strong 
magnetic field normal to the triangular plane. We further evaluate the 
effective-spin correlation within the mean-field spherical approximation. 
We explicitly demonstrate that, the nearest-neighbor anisotropic effective-spin 
interaction, originating from the strong spin-orbit coupling, 
enhances the effective-spin correlation at the M points in the Brillouin zone. 
We identify these results as the strong evidence for the anisotropic 
interaction and the strong spin-orbit coupling in YbMgGaO$_4$.
\end{abstract}

\maketitle

\section{Introduction}
\label{sec1}

The rare earth triangular lattice antiferromagnet 
YbMgGaO$_4$ was recently proposed to be a candidate for quantum spin liquid 
(QSL)~\cite{YueshengPRL,YaodongPRB,ShenYao201607,Martin201607,YueshengmuSR}. 
In YbMgGaO$_4$, the Yb$^{3+}$ ions form a perfect 
two-dimensional triangular lattice.  
For the Yb$^{3+}$ ions, the strong spin-orbit coupling (SOC)
entangles the orbital angular momentum, ${\bf L}$ ($L=3$),
with the total spin, ${\bf s}$ ($s=1/2$), leading to 
a total moment, ${\bf J}$ ($J=7/2$)~\cite{YueshengPRL,YaodongPRB}. 
Like the case in the spin ice material Yb$_2$Ti$_2$O$_7$~\cite{Ross2011},
the crystal electric field in YbMgGaO$_4$ further splits the 
eight-fold degeneracy of the Yb$^{3+}$ total moment into  
four Kramers' doublets. The ground state Kramers' doublet 
is separated from the excited doublets by a crystal field 
energy gap. At the temperature that is much lower than the 
crystal field gap, the magnetic properties of YbMgGaO$_4$ 
are fully described by the ground state Kramers' doublets~\cite{YaodongPRB}. 
The ground state Kramers' doublet is modeled by an 
effective-spin-1/2 local moment ${\bf S}$. Therefore, 
YbMgGaO$_4$ is regarded as a QSL with effective-spin-1/2 
local moments on a triangular lattice~\cite{YueshengPRL,YaodongPRB,ShenYao201607,Martin201607}.

The existing experiments on YbMgGaO$_4$ have involved 
 thermodynamic, neutron scattering, and $\mu$SR measurements~\cite{YueshengPRL,ShenYao201607,YueshengmuSR,Martin201607}. 
The system was found to remain disordered down to 0.05K in the 
recent $\mu$SR measurement~\cite{YueshengmuSR}. 
The thermodynamic measurement finds a constant magnetic susceptibility 
in the zero temperature limit. In the low temperature regime, 
the heat capacity~\cite{YueshengPRL,Martin201607}  
behaves as $C_v \approx \text{constant} \times T^{0.7}$. 
The inelastic neutron scattering measurements from 
two research groups have found the presence of broad 
magnetic excitation continuum~\cite{ShenYao201607,Martin201607}. 
In particular, the inelastic neutron scattering results from 
Yao Shen {\it et al} clearly indicate the upper excitation edge and 
the {\sl dispersive} continuum of magnetic excitations~\cite{ShenYao201607}.  
Both neutron scattering results found a weak spectral peak at the M points 
in the Brillouin zone~\cite{ShenYao201607,Martin201607}. 
Based on the existing experiments, we have proposed that 
the spinon Fermi surface U(1) QSL gives a reasonable description 
of the experimental results~\cite{ShenYao201607}.

%-----------------------------
\begin{figure}[t]
\centering
{\includegraphics[width=6.5cm]{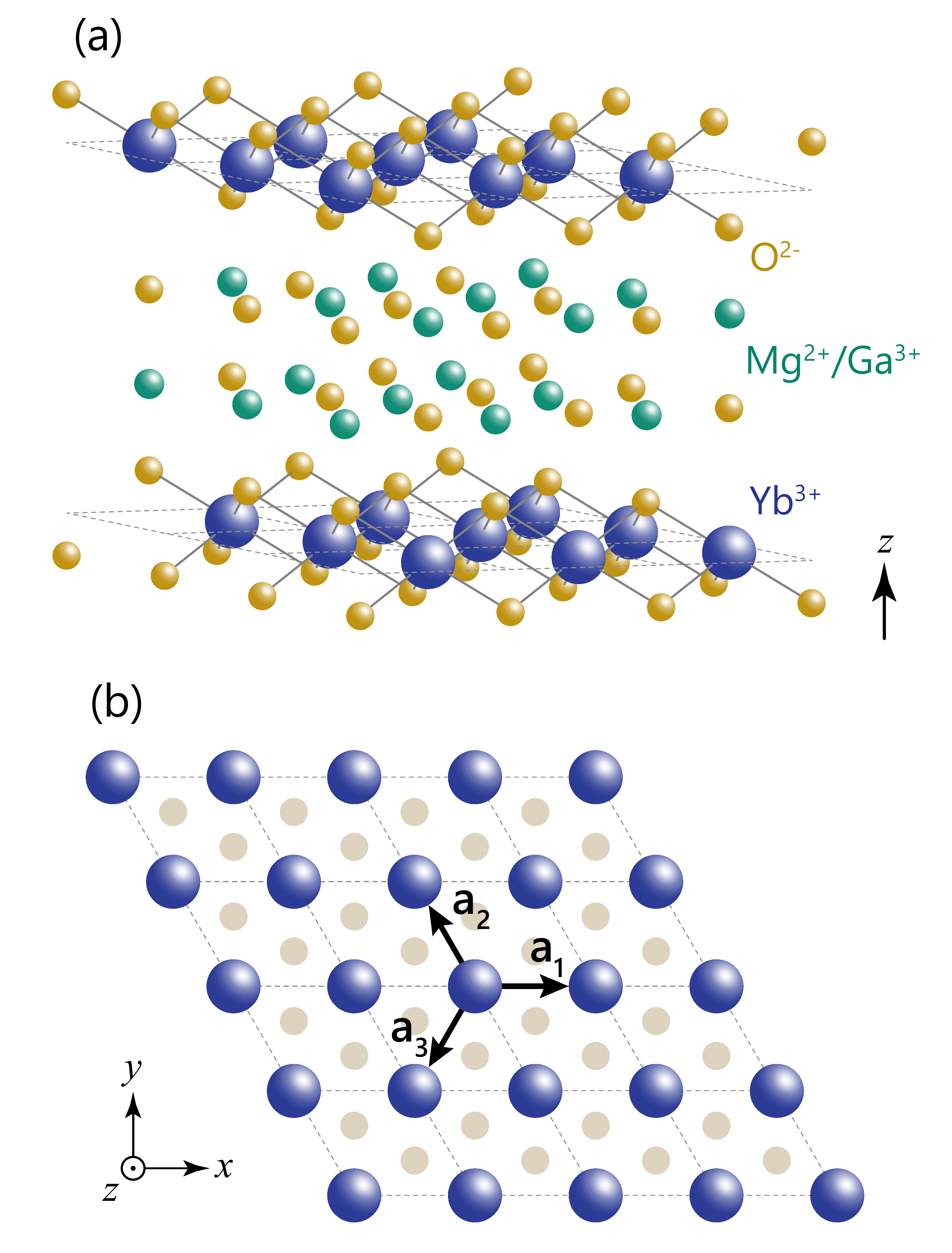}}
 \caption{(Color online.)
(a) The crystal structure of YbMgGaO$_4$. Mg and Ga ions form the non-magnetic layer. 
(b) The Yb triangular layer. }
\label{fig1}
\end{figure}
%-----------------------------

Previously, two organic triangular antiferromagnets, 
$\kappa$-(ET)$_2$Cu$_2$(CN)$_3$ and EtMe$_3$Sb[Pd(dmit)$_2$]$_2$, 
were proposed to be QSLs~\cite{kappaET,dmit,organics1,organics2}. 
These two materials are in the weak Mott regime, 
where the charge fluctuation is strong. 
It was then suggested that the four-spin ring exchange interaction
due to the strong charge fluctuation may destabilize the 
magnetic order and favor a QSL ground state~\cite{Motrunich2005,Lee05}.  
Unlike the organic counterparts, YbMgGaO$_4$ is in the strong Mott 
regime~\citep{YaodongPRB,YueshengPRL}. 
The $4f$ electrons of the Yb$^{3+}$ ion is very localized spatially.
As a result, the physical mechanism for the QSL ground state 
in this new material is deemed to be quite different. 
The new ingredient of the new material is the strong SOC and  
the spin-orbit entangled nature of the Yb$^{3+}$ local moment.
It was pointed out that the spin-orbit entanglement 
leads to highly anisotropic interactions between the
Yb local moments~\cite{YaodongPRB,WCKB,Chen2008,PhysRevB.82.174440}. The anisotropic 
effective-spin interaction is shown to enhance the 
quantum fluctuation and suppress the magnetic order 
in a large parameter regime where the QSL may be located~\cite{YaodongPRB}. 
On the fundamental side, it was recently argued that, 
as long as the time reversal symmetry 
is preserved, the ground state of a spin-orbit-coupled Mott 
insulator with odd number of electrons per cell must be 
exotic~\cite{watanabe2015filling}.
This theoretical argument implies that the 
spin-orbit-coupled Mott insulator can in principle be 
candidates for spin liquids. YbMgGaO$_4$ falls into this class 
and is actually the first such material.

More recently, Ref.~\onlinecite{Martin201607} introduced 
the XXZ exchange interactions on both nearest-neighbor and next-nearest-neighbor 
sites to account for the spin-wave dispersion in the strong 
magnetic field and the weak peak at the M points in the 
effective-spin correlations. The authors further suggested the 
further neighbor competing exchange interactions as the possible mechanism 
for the QSL in YbMgGaO$_4$. In this paper, however, we focus on 
the anisotropic effective-spin interactions on the nearest-neighbor 
sites. After carefully justifying the underlying microscopics that 
supports the nearest-neighbor anisotropic model, we demonstrate 
that the nearest-neighbor model is sufficient to reproduce the 
spin-wave dispersion of the polarized state in the strong magnetic field. 
With the nearest-neighbor anisotropic model, we further show that the 
effective-spin correlation also develops a peak at the M points. 
Therefore, we think the nearest-neighbor anisotropic 
model captures the essential physics for YbMgGaO$_4$.

The remaining part of the paper is outlined as follows. 
In Sec.~\ref{sec2}, we describe some of the details about the 
microscopics of the interactions between the Yb local 
moments. In Sec.~\ref{sec3}, we compare the spin-wave dispersion 
of the nearest-neighbor anisotropic interactions in a strong 
field with the existing experimental data. In Sec.~\ref{sec4}, 
we evaluate the effective-spin correlation from the effective-spin 
models with and without the anisotropic interaction. 
Finally in Sec.~\ref{sec5}, we conclude with a discussion. 

%-----------------------------
\begin{figure*}[ht]
\centering
{\includegraphics[width=\textwidth]{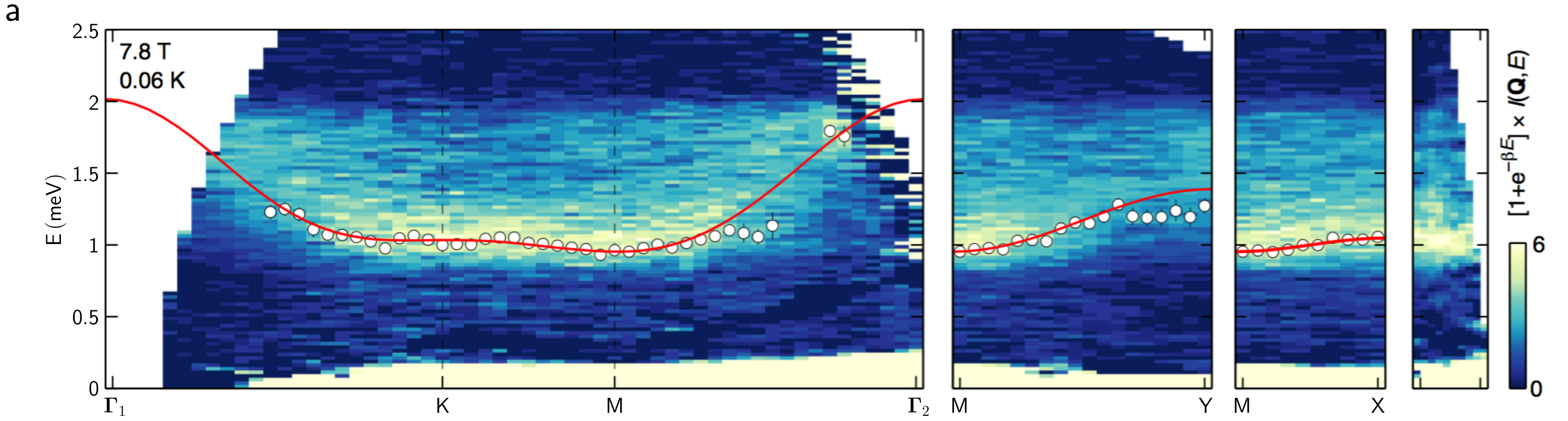}}
\centering
{\includegraphics[width=\textwidth]{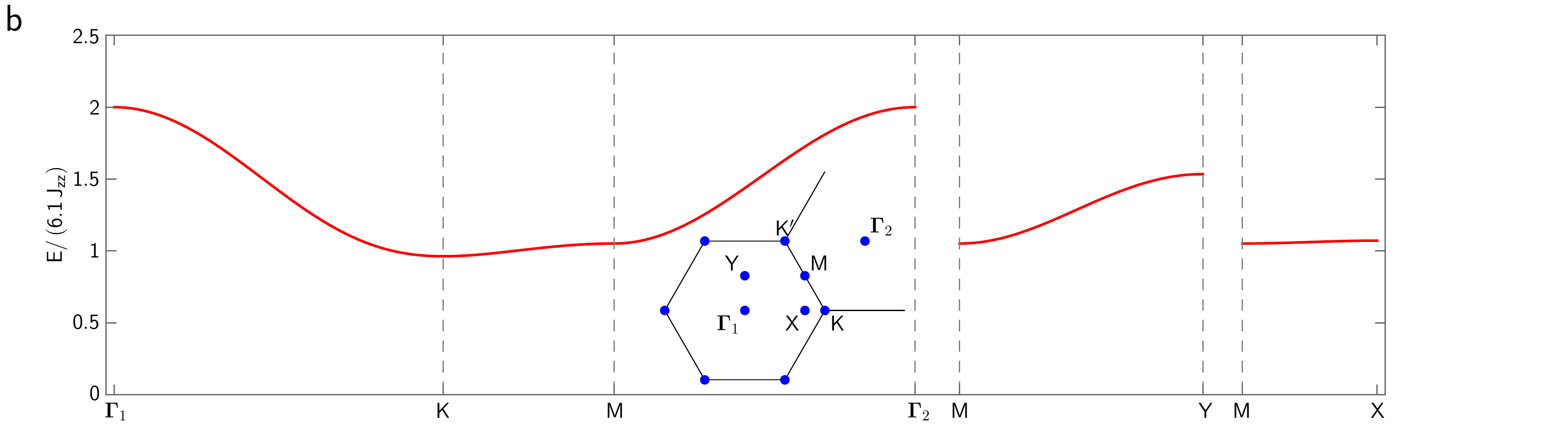}}
\caption{(Color online.)
(a) The experimental spin-wave dispersion in the presence of external 
field along $z$-direction with 
field strength 7.8T at 0.06K (adapted from Ref.~\onlinecite{Martin201607}). According to Ref.~\onlinecite{Martin201607}, 
the white circles indicate the location of the maximum intensity. 
The error bar, however, was not indicated in the plot. 
The red lines show a fit to the spin-wave dispersion relation 
that is obtained after including both nearest-neighbor and 
next-nearest-neighbor XXZ exchange interactions~\cite{Martin201607}. 
(b) The theoretical spin-wave dispersion according to
the nearest-neighbor anisotropic exchange model Eq.~(\ref{eq1}), 
where we set $J_{\pm}/J_{zz} = 0.66, J_{\pm\pm}/J_{zz}=0.34$, 
and $h/J_{zz} = 10.5$. The analytical expression of 
the dispersion is given in Eq.~(\ref{eq2}).
The inset of (b) is the Brillouin zone. 
}
\label{fig2}
\end{figure*}
%-----------------------------

\section{The anisotropic interaction for the effective spins}
\label{sec2}

Compared to the organic spin liquid candidates~\cite{kappaET,dmit,organics1,organics2}, 
YbMgGaO$_4$ is in the strong Mott regime, and the charge fluctuation is rather weak. 
Therefore, the four-spin ring exchange, that is a higher order       
perturbative process than the nearest-neighbor pairwise interaction, 
is strongly suppressed. In the previous work~\citep{YaodongPRB,YueshengPRL}, 
we have proposed that the generic pairwise effective-spin interaction 
for the nearest-neighbor Yb moments in YbMgGaO$_4$,
\begin{eqnarray}
{\mathcal H} &=& \sum_{\langle{{\bf r}{\bf r}'}\rangle}
J_{zz}^{} S_{\bf r}^z S_{{\bf r}'}^z 
+ J_{\pm}^{} ( S_{\bf r}^+ S_{{\bf r}'}^- 
+ S_{\bf r}^- S_{{\bf r}'}^+ )
\nonumber \\
&& + J_{\pm\pm}^{} (  
\gamma_{{\bf r}{\bf r}' }^{} S_{{\bf r}}^+ S_{{\bf r}'}^+ 
+  \gamma_{ {\bf r}{\bf r}'  }^{\ast} S_{\bf r}^- S_{{\bf r}'}^- )
	\nonumber \\
&& - \frac{i J_{z\pm}}{2}
    \big[ (\gamma_{{\bf r}{\bf r}' }^{\ast} S_{\bf r}^+ 
    - \gamma_{ {\bf r}{\bf r}'}^{} S_{\bf r}^- ) S_{{\bf r}'}^z
    \nonumber \\
&& \quad\quad\quad
     +  S_{\bf r}^z  (\gamma_{ {\bf r}{\bf r}'}^{\ast} S_{{\bf r}'}^+ 
     - \gamma_{ {\bf r}{\bf r}'}^{} S_{{\bf r}'}^- )  \big],
\label{eq1}
\end{eqnarray}
where $S_{\bf r}^\pm = S_{\bf r}^x \pm i S_{\bf r}^y$, 
and $\gamma_{{\bf r}{\bf r}'} =\gamma_{{\bf r}'{\bf r}} 
= 1, e^{i2\pi/3}, e^{-i2\pi/3}$ are the phase factors 
for the bond ${\bf r}{\bf r}'$ along the \textbf{a$_1$}, 
\textbf{a$_2$}, \textbf{a$_3$} directions (see Fig.~\ref{fig1}).
The $J_{\pm\pm}$ and $J_{z\pm}$ terms of Eq.~(\ref{eq1}) 
are anisotropic interactions arising naturally from the strong SOC. 
Due to the SOC, the effective-spins inherit the symmetry 
operation of the space group, hence there are bond-dependent
$J_{\pm\pm}$ and $J_{z\pm}$ interactions.

Our generic model in Eq.~(\ref{eq1}) contains the contribution from
all microscopic processes that include the direct $4f$-electron exchange, 
the indirect exchange through the intermediate oxygen ions, and 
the dipole-dipole interaction. The further neighbor interaction is 
neglected in our generic model. Like the ring exchange, 
the further neighbor superexchange usually involves higher order 
perturbative processes via multiple steps of electron tunnelings 
than the nearest-neighbor interactions. Even though the further neighbor 
superexchange interaction can be mediated by the direct electron hoppings 
between these sites, the contribution should be very small due to the very
localized nature of the $4f$ electron wavefunction. 
The remaining contribution is the further neighbor dipole-dipole interaction. 
For the next-nearest neighbors, the dipole-dipole interaction
is estimated to be $\sim$0.01-0.02K and is thus one or two orders of magnitude
smaller than the nearest-neighbor interactions. Therefore, we can safely 
neglect the further neighbor interactions and only keep the nearest 
neighbor ones in Eq.~(\ref{eq1}).

The large chemical difference prohibits the Ga or Mg 
contamination in the Yb layers. 
The Yb layers are kept clean, and there is little
disorder in the exchange interaction. Although there exists Ga/Mg 
mixing in the nonmagnetic layers, the exchange path that they 
involve would be Yb-O-Ga-O-Yb or Yb-O-Mg-O-Yb (see Fig.~\ref{fig1}). 
This exchange path is a higher order perturbative process     
than the Yb-O-Yb one and thus can be neglected. 
We do not expect the Ga/Mg mixing in the nonmagnetic layers 
to cause much exchange disorder within the Yb layers. 
The Ga/Mg disorder in YbMgGaO$_4$ is different from the 
Cu/Zn disorder in herbertsmithite~\cite{kagome_NMR,Han2012,fu2015evidence,han2015correlated}. 
In the latter case, the Cu disorder carries magnetic moment 
and directly couples to the spin in the Cu layers. 

The XXZ limit of our generic model has already been studied 
in some of the early works~\cite{XXZ2014,PhysRevB.91.081104}. 
It was shown that the magnetic ordered 
ground state was obtained for all parameter regions in the XXZ
limit. To obtain a disordered ground state for the generic model,
it is necessary to have the $J_{\pm\pm}$ and $J_{z\pm}$ 
interactions. In Ref.~\onlinecite{YaodongPRB}, we have shown 
that the 120-degree magnetic order in the XXZ limit is actually 
destabilized by the enhanced quantum fluctuation 
when the anisotropic $J_{\pm\pm}$ and $J_{z\pm}$ 
interactions are introduced.

\section{Spin-wave dispersion in the strong magnetic field}
\label{sec3}

The nearest-neighbor interaction between the Yb local moments are of 
the order of several Kelvins~\cite{YueshengPRL}; as a result, 
a moderate magnetic field in the lab is sufficient for polarizing the 
local moment~\cite{YaodongPRB}. Under the linear spin-wave 
approximation, the spin-wave dispersion in the presence of the 
strong external magnetic field is given as~\cite{YaodongPRB}
\begin{eqnarray}
\omega_{z} ({{\bf k}}) & = & 
\Big\{ \big[  g_z \mu_B B_z - 3 J_{zz}
+ 2 J_{\pm} \sum_{i=1}^3 \cos ( {\bf k}\cdot {\bf a}_i) \big]^2
\nonumber \\
&& \quad 
- 4 J_{\pm\pm}^2 \big| \cos ( {\bf k}\cdot {\bf a}_1)
+ e^{-i \frac{2\pi}{3} }\cos ( {\bf k}\cdot {\bf a}_2)
\nonumber \\
&&
\quad 
+ e^{i \frac{2\pi}{3} } \cos ( {\bf k} \cdot {\bf a}_3) 
\big|^2 \Big\}^{ 1/2 },
\label{eq2}
\end{eqnarray}
where $g_z$ and $B_z$ are Land\'{e} factor and magnetic field along $z$-direction, 
respectively. Notice that the dispersion in Eq.~(\ref{eq2}) is independent of 
$J_{z\pm}$; this is an artifact of the {\sl linear} spin-wave approximation.

In the recent experiment in Ref.~\onlinecite{Martin201607},   
a magnetic field of 7.8T normal to the Yb plane at 0.06K,
a gapped magnon band structure is observed. In Fig.~\ref{fig2}, 
we compare our theoretical result with a tentative choice of 
exchange couplings in Eq.~(\ref{eq2}) with the experimental 
results from Ref.~\onlinecite{Martin201607}. Since the error bar 
is not known from Ref.~\onlinecite{Martin201607}, judging from 
the extension of the bright region in Fig.~\ref{fig2}a, 
we would think that the agreement between the theoretical result and
the experimental result is reasonable. Here, we have to mention that  
the dispersion that is plotted in Fig.~\ref{fig2} is not quite 
sensitive to the choice of $J_{\pm\pm}$. Therefore, we expect
it is better to combine the spin-wave dispersion for several 
field orientations and to extract the exchange couplings 
more accurately. For an arbitrary external field in the $xz$ plane, 
the Hamiltonian is given by
\begin{eqnarray}
{\mathcal H}_{xz}^{} = {\mathcal H} 
- \sum_{\bf r} \mu_B^{} \big[ g_x^{} B_x^{} S_{\bf r}^x 
+ g_z^{} B_z^{} S_{\bf r}^z \big].
\end{eqnarray}
Since $g_x \neq g_z$, the uniform magnetization,
${\bf m} \equiv \langle {\bf S}_{\bf r} \rangle$,
is generally not parallel to the external magnetic field. 
For $B_x \equiv B \sin \theta$ and $B_z \equiv  B \cos \theta $,
the magnetization is given by 
\begin{equation}
{\bf m} = m (\hat{x} \sin \theta' 
+ \hat{z} \cos \theta' ),
\end{equation} 
where $\tan \theta' = (g_x/g_z) \tan \theta $.
At a sufficiently large magnetic field, all the moments are polarized along 
the direction defined by $\theta'$.  
In the linear spin-wave theory for this polarized state, we choose the magnetization to be 
the quantization axis for the Holstein-Primakoff transformation, 
\begin{eqnarray}
&& {\bf S}_{\bf r} \cdot \frac{\bf m}{|{\bf m}|} \equiv \frac{1}{2}-a^{\dagger}_{\bf r} a^{}_{\bf r}
,
\\
&& {\bf S}_{\bf r} \cdot \hat{y} \equiv  \frac{1}{2}(a_{\bf r}^{} + a^\dagger_{\bf r} ) ,
\\
&& {\bf S}_{\bf r} \cdot ( \frac{\bf m}{|{\bf m}|} \times \hat{y})
 \equiv 
 \frac{1}{2i}(a_{\bf r}^{} - a^\dagger_{\bf r} ) ,
\end{eqnarray}
where $a_{\bf r}^{\dagger}$ ($a_{\bf r}^{}$) is the creation 
(annihilation) operator for the Holstein-Primakoff boson. 
In the linear spin-wave approximation, we plug the Holstein-Primakoff
transformation into ${\mathcal H}_{xz}$ and keep the quadratic part of the 
the Holstein-Primakoff bosons. The spin-wave dispersion is obtained 
by solving the linear spin-wave Hamiltonian and is given by 

\begin{widetext}

\begin{eqnarray}
    \omega_{xz} ({\bf k}) &&=
    \Big\{
        \Big[ g_x \mu_B^{} B_x \sin \theta'
            +g_z \mu_B^{} B_z \cos \theta'
            -6J_{\pm} \sin^2 \theta'
            - {3J_{zz} } \cos^2 \theta' 
            + \cos({\bf k}\cdot{\bf a}_1)
                \Big(
                     \frac{J_{\pm}}{2} (3 + \cos 2\theta') 
            \nonumber \\
            -&&
                     {J_{\pm\pm}} \sin^2 \theta'
                    +\frac{J_{zz}}{2} \sin^2 \theta'
                \Big)
                +\cos({\bf k}\cdot{\bf a}_2)
                \Big(
                     \frac{J_{\pm}}{2} (3 + \cos 2\theta' )
                      +\frac{J_{\pm\pm}}{2} \sin^2 \theta'
                      +\frac{J_{zz}}{2} \sin^2 \theta'
            \nonumber \\
                  +&& \frac{\sqrt{3}}{4} J_{z\pm} \sin 2\theta'
                \Big)
           +\cos({\bf k}\cdot{\bf a}_3)
                \Big(
                     \frac{J_{\pm}}{2} (3 + \cos 2\theta' )
                    +\frac{J_{\pm\pm}}{2}\sin^2 \theta'
                    +\frac{J_{zz}}{2} \sin^2 \theta'
                    -\frac{\sqrt{3}}{4} J_{z\pm} \sin 2\theta'
                \Big)
        \Big]^2
        \nonumber \\
        -&&
        \Big|
            \cos({\bf k}\cdot{\bf a}_1)
                \Big(
                    J_\pm \sin^2 \theta'
                    -J_{\pm\pm}(1+\cos^2\theta' )
                    -iJ_{z\pm}\sin \theta'
                    -\frac{J_{zz}}{2}\sin^2\theta'
                \Big)
            \nonumber \\
            +&& \cos({\bf k}\cdot{\bf a}_2)
                \Big(
                    {J_{\pm}}\sin^2 \theta'
                    +\frac{J_{\pm\pm}}{4} (3 + \cos 2\theta' - 4 i \sqrt{3} \cos \theta' )
                    -\frac{J_{zz}}{2} \sin^2 \theta'
                    +\frac{J_{z\pm}}{4} (2i\sin\theta' - \sqrt{3} \sin2\theta' )
                \Big)
            \nonumber \\
            +&&\cos({\bf k}\cdot{\bf a}_3)
                \Big(
                    {J_{\pm}} \sin^2 \theta'
                    +\frac{J_{\pm\pm}}{4} (3 + \cos 2\theta' + 4 i \sqrt{3} \cos \theta')
                    -\frac{J_{zz}}{2} \sin^2 \theta'
                    +\frac{J_{z\pm}}{4} (2i\sin\theta' + \sqrt{3} \sin2\theta' )
                \Big)
        \Big|^2 \Big\}^{1/2}.
\end{eqnarray}

Likewise, for the field within the $xy$ plane, the Hamiltonian is given by
\begin{eqnarray}
{\mathcal H}_{xy}^{} = {\mathcal H} 
- \sum_{\bf r} \mu_{B}^{}\big[ g_x^{} B_x^{} S_{\bf r}^x 
+ g_y^{} B_y^{} S_{\bf r}^y \big].
\end{eqnarray}
Now because of the three-fold on-site symmetry, $g_x = g_y$.
The magnetization is parallel to the external magnetic field.  
For $B_x \equiv B \cos \phi$ and $B_y \equiv  B \sin \phi$, 
the magnetization is ${\bf m} = m (\hat{x} \cos \phi
+ \hat{y} \sin \phi )$, and the corresponding spin-wave dispersion
in the strong field limit is given by
\begin{eqnarray}
    \omega_{xy} ({\bf k}) && =
    \Big\{
        \Big[ g_x \mu_B B_x \cos\phi
            +g_y \mu_B B_y \sin\phi
            -6J_{\pm}
            +\cos({\bf k}\cdot{\bf a}_1)
                \big(
                     J_{\pm}
                    +\frac{J_{zz}}{2}
                    - J_{\pm\pm} \cos 2\phi
                \big)
            \nonumber \\
            + && \cos({\bf k}\cdot{\bf a}_2)
                \Big(
                     J_{\pm}
                    +\frac{J_{zz}}{2}
                    +{J_{\pm\pm}} \cos (2\phi-\frac{\pi}{3}) 
                \Big)
                +\cos({\bf k}\cdot{\bf a}_3)
                \Big(
                     J_{\pm}
                    +\frac{J_{zz}}{2}
                    +{J_{\pm\pm}} \cos (2\phi+\frac{\pi}{3}) 
                \Big)
        \Big]^2
            \nonumber \\
        -   \Big|&&
            \cos({\bf k}\cdot{\bf a}_1)
                \Big(
                     J_{\pm}
                    - \frac{J_{zz}}{2}
                    - J_{\pm\pm} \cos 2\phi
                    + i J_{z\pm} \cos \phi
                \Big)+ \cos({\bf k}\cdot{\bf a}_2)
                \Big(
                     J_{\pm}
                    - \frac{J_{zz}}{2}
                    + {J_{\pm\pm}}\cos (2\phi + \frac{\pi}{3})
            \nonumber \\
            -&&
                    {i J_{z\pm}} \cos( \phi- \frac{\pi}{3})
                \Big)
            + \cos({\bf k}\cdot{\bf a}_3)
                \Big(
                     J_{\pm}
                    - \frac{J_{zz}}{2}
                    + {J_{\pm\pm}} \cos (2\phi - \frac{\pi}{3})
                    - {i J_{z\pm}} \cos (\phi+ \frac{\pi}{3})
                \Big)
        \Big|^2    \Big\}^{1/2}.
\end{eqnarray}
\end{widetext}

%-----------------------------
\begin{figure}[ht]
\centering
{\includegraphics[height=.26\textwidth]{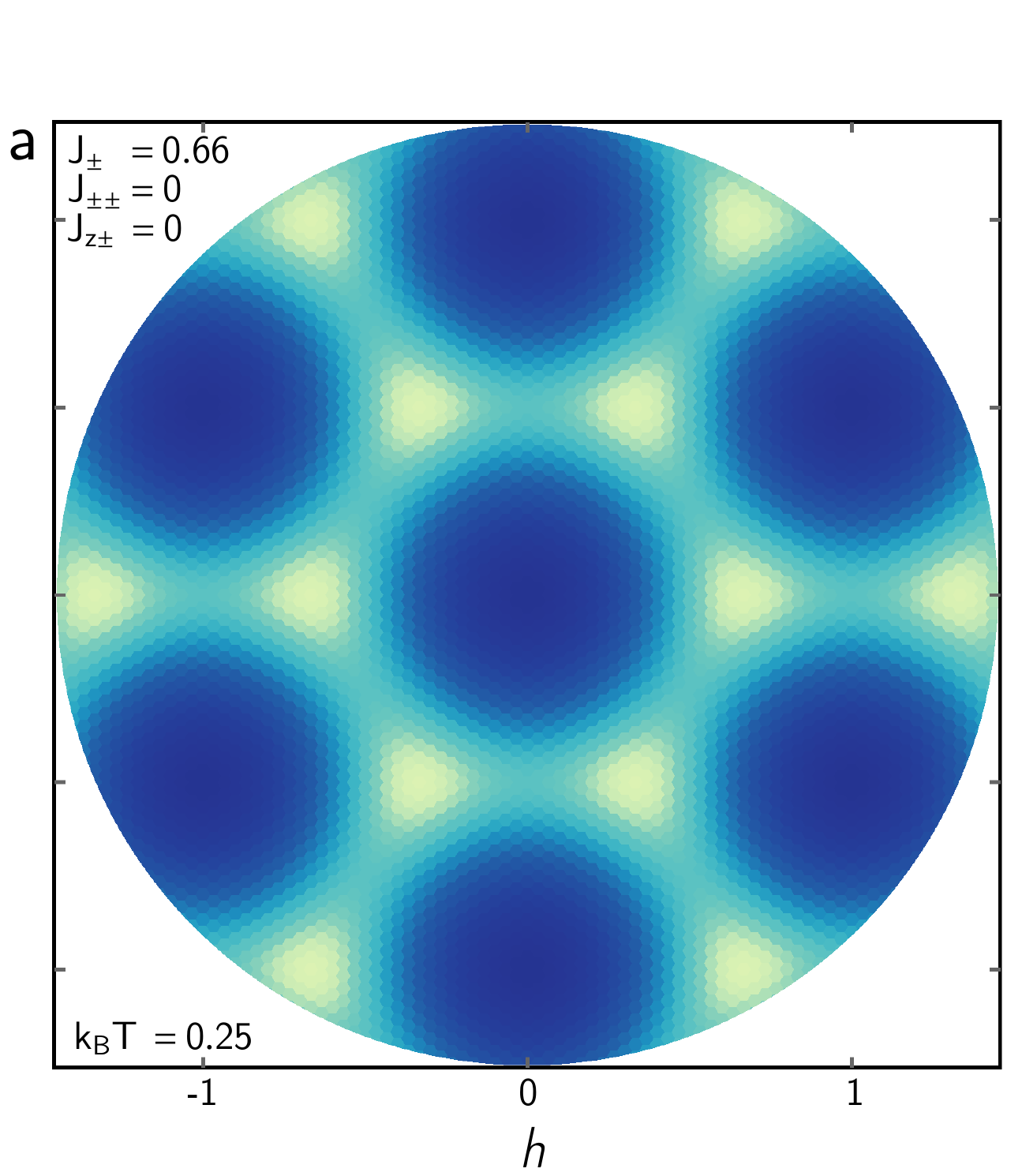}
 \includegraphics[height=.26\textwidth]{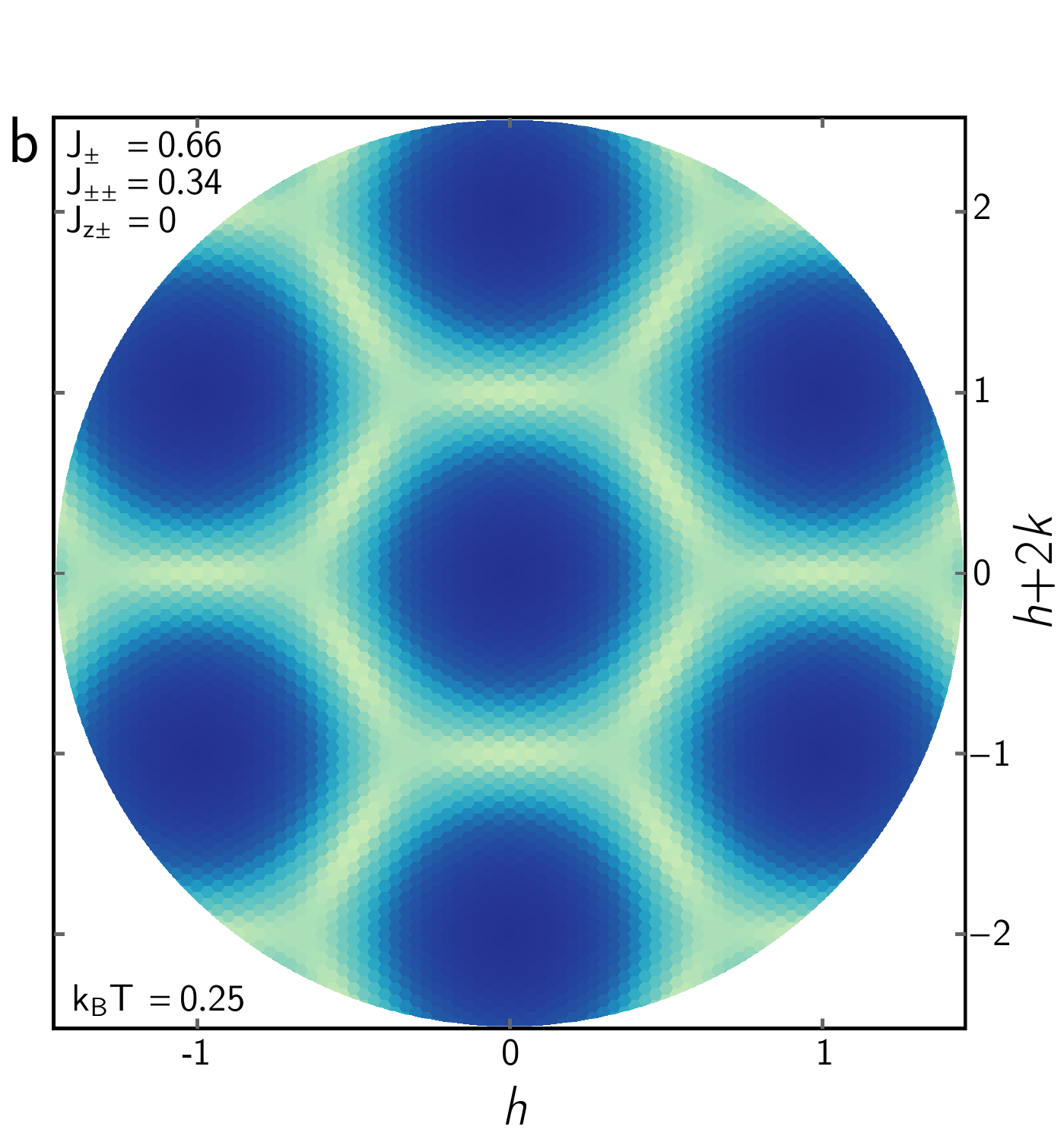}
 \includegraphics[height=.26\textwidth]{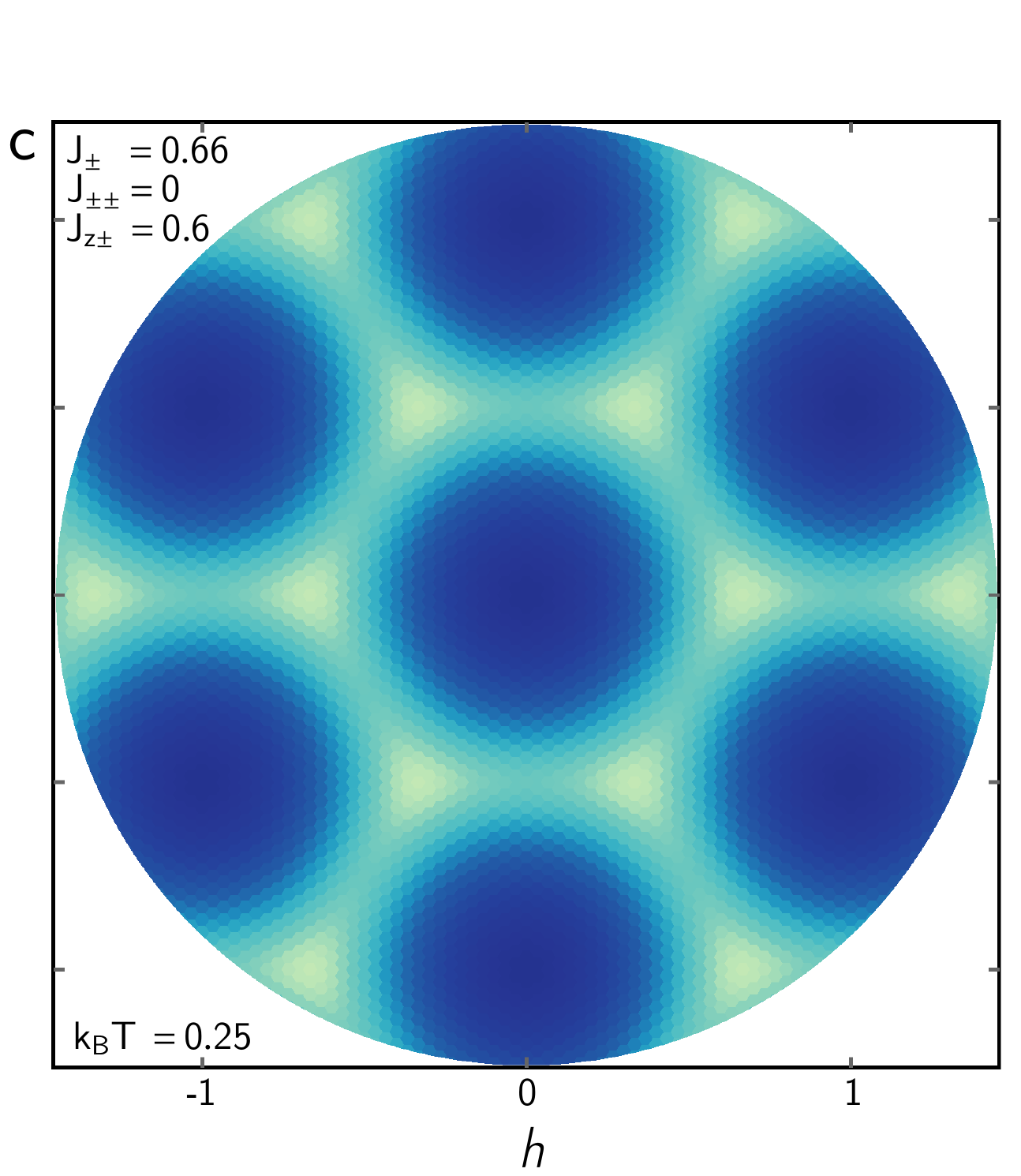}
 \includegraphics[height=.26\textwidth]{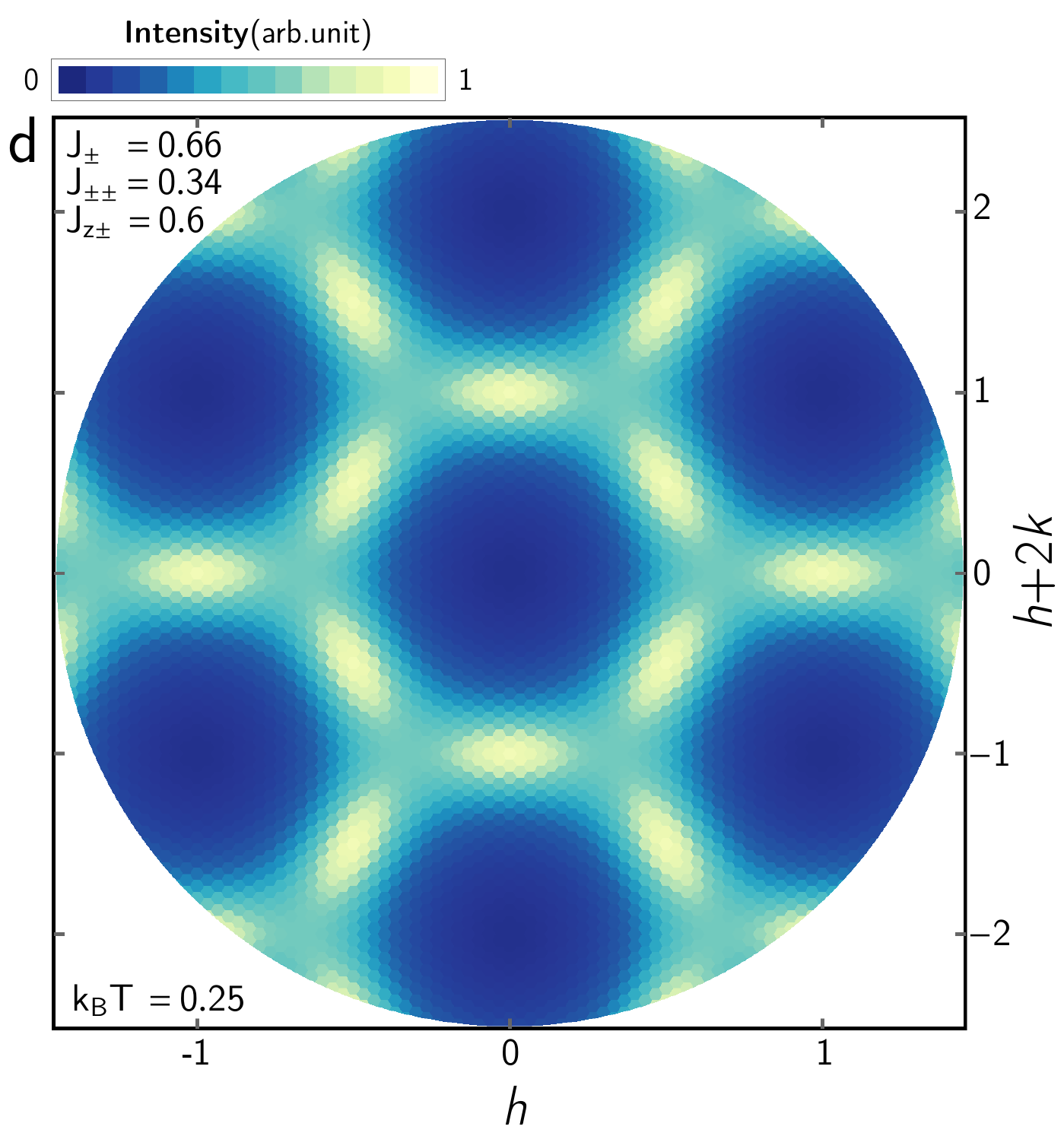}}
\caption{(Color online.) Contour plot of effective-spin correlation  
$\avg{S^+_{\bf k}\, S^-_{\bf -k}}$ in the momentum space. 
The correlation function is computed from the nearest-neighbor model 
in Eq.~(\ref{eq1}), with parameters in units of $J_{zz}$ indicated.
Without the anisotropic exchanges, the spectral weight peaks around K. 
The anisotropic $J_{\pm\pm}$ and $J_{z\pm}$ interactions can switch the peak to 
M.
}
%-----------------------------

\label{fig3}
\end{figure}

\section{Effective-spin correlation}
\label{sec4}

In both Ref.~\onlinecite{ShenYao201607} 
and Ref.~\onlinecite{Martin201607}, a weak spectral peak at the M points 
is found in the inelastic neutron scattering data. This result 
indicates that the interaction between the Yb local moments 
enhances the correlation of the effective-spins at the M points.
Actually in Ref.~\onlinecite{YaodongPRB}, we have already shown that, 
the anisotropic $J_{\pm\pm}$ and $J_{z\pm}$ interactions, if they are significant,
would favor a stripe magnetic order with an ordering wavevector at the M points~\cite{Rongyu}.
This theoretical result immediately means that the anisotropic $J_{\pm\pm}$ 
and $J_{z\pm}$ interactions would enhance the effective-spin correlation
at the M points. In the following, we demonstrate explicitly that the 
generic model in Eq.~(\ref{eq1}) with the anisotropic nearest-neighbor 
interactions does enhance the effective-spin correlation at the M points. 
We start from the mean-field partition function of the system, 
\begin{eqnarray}
	\mathcal{Z}
	&=&  \int \mathcal{D}  \left[ \bf{S}_{\bf r} \right] 
	\prod_{\bf r} \delta({\bf S}_{\bf r}^2 - S^2) \, 
	e^{-\beta \mathcal{H}}  \nonumber \\
	&=&  \int \mathcal{D} \left[ \bf{S}_{\bf r} \right]
	 \mathcal{D} [\lambda_{\bf r} ] \,
	 e^{-\beta \mathcal{H} +  \sum_{\bf r} \lambda_{\bf r} [ {\bf S}_{\bf r}^2 - S^2 ] } 
	 \nonumber \\
	&\equiv& \int \mathcal{D} \left[ \bf{S}_{\bf r} \right]
	  \mathcal{D} [\lambda_{\bf r} ] \,   
	  e^{-\mathcal{S}_{\text{eff}} [\beta, \lambda_{\bf r} ] },
\end{eqnarray}
where $\mathcal{H}$ is given in Eq.~(\ref{eq1}), 
$\mathcal{S}_{\text{eff}}$ is the effective action 
that describes the effective-spin interaction, and 
$\lambda_{\bf r}$ is the local Lagrange multiplier 
that imposes the local constraint with $|{\bf S}_{\bf r}|^2 = S^2$. 
Although this mean-field approximation does not 
gives the quantum ground state, it does provide
a qualitative understanding about the relationship
between the effective-spin correlation 
and the microscopic spin interactions. 

To evaluate the effective-spin correlation, we here adopt 
a spherical approximation~\cite{bergman2007order} by replacing 
the local constraint with a global one such that 
$\sum_{\bf r} |{\bf S}_{\bf r}|^2 = N_{\text{site}} S^2 $,
where $N_{\text{site}}$ is the total number of lattice sites.
This approximation is equivalent to choosing a uniform 
Lagrange multiplier with $\lambda_{\bf r} \equiv \lambda$. 
It has been shown that the spin correlations determined 
from classical Monte Carlo simulation are described 
quantitatively within this scheme~\cite{bergman2007order}. 

In the momentum space, we define 
\begin{equation}
S^{\mu}_{\bf r} \equiv \frac{1}{\sqrt{N_{\text{site}}}} 
\sum_{{\bf k}\in \text{BZ}} S^{\mu}_{\bf k} \, e^{i {\bf k} \cdot {\bf r}},
\end{equation} 
and the effective action is given by
\begin{eqnarray}
\mathcal{S}_{\text{eff}} [ \beta, \lambda ]  
&=& \sum_{{\bf k} \in {\rm BZ}} 
\beta \big[ {\mathcal J}_{\mu \nu}^{}({\bf k}) + \Delta(\beta) \delta_{\mu\nu}^{} \big]
S^\mu_{\bf k}\, S^\nu_{{\bf -k}} 
\nonumber \\
&& \quad\quad - \beta N_{\text{site}} \Delta (\beta) S^2,
\end{eqnarray}
where we have placed $\lambda \equiv -\beta \Delta (\beta)$ in a 
saddle point approximation, $\mu, \nu = x, y, z$, and ${\mathcal J}_{\mu \nu}({\bf k})$ 
is a 3$\times$3 exchange matrix that is obtained from Fourier transforming the 
exchange couplings. 
Note the XXZ part of the spin interactions only appears in the 
diagonal part of ${\mathcal J}_{\mu \nu}({\bf k})$ while
the anisotropic $J_{\pm\pm}$ and $J_{z\pm}$ interactions 
are also present in the off-diagonal part. 
Hence the effective-spin correlation is given as 
\begin{equation}
\avg{S^\mu_{\bf k}\, S^\nu_{\bf -k}} = \frac{1}{\beta}
\left[ {\mathcal J}({\bf k})^{} + \Delta (\beta) \mathbb{1}_{3\times 3}^{}
 \right]_{\mu \nu}^{-1},
 \label{eq14}
\end{equation}
where $\mathbb{1}_{3\times 3}$ is a 3$\times$3 identity matrix. 

The saddle point equation is obtained by integrating out the effective-spins
in the partition function and is given by
\begin{equation}
 \sum_{{\bf k} \in {\rm BZ}}
\sum_{\mu} \frac{1}{\beta}
\left[ {\mathcal J}({\bf k}) + \Delta (\beta) \mathbb{1}_{3\times 3}
 \right]_{\mu \mu}^{-1} = N_{\rm site} S^2,
\end{equation}
from which, we determine $\Delta (\beta)$ and the effective-spin correlation
in Eq.~(\ref{eq14}). 

The results of the effective-spin correlations are presented in 
Fig.~\ref{fig3}. In the absense of the $J_{\pm\pm}$ and $J_{z\pm}$ 
interactions, the correlation function is peaked at the K points.
This result is understood since the XXZ model favors the 120-degree state
would simply enhance the effective-spin correlation at the K points that
correspond to the ordering wavevectors of the 120-degree state. 
After we include the $J_{\pm\pm}$ 
and $J_{z\pm}$ interactions, the peak of the correlation function 
is switched to the M points (see Fig.~\ref{fig3}d). 
This suggests that it is sufficient to have the $J_{\pm\pm}$
and $J_{z\pm}$ interactions in the nearest-neighbor model 
to account for the peak at the M points 
in the neutron scattering results.

\section{Discussion}
\label{sec5}

Instead of invoking further neighbor interaction 
in Ref.~\onlinecite{Martin201607}, we have focused on the 
anisotropic spin interaction on the nearest-neighbor bonds to 
account for the spin-wave dispersion of the polarized state 
in the strong magnetic field and the effective-spin 
correlation in YbMgGaO$_4$. The bond-dependent interaction 
is a natural and primary consequence of the strong SOC in the system.  
As for the importance of further neighbor interaction, 
it might be possible that some physical mechanism,
that we are not aware of,
may suppress the anisotropic spin interactions between the 
nearest neighbors but enhance the further neighbor spin interactions.

We have recently proposed that the spinon Fermi surface U(1) QSL 
provides a consistent explanation for the experimental 
results in YbMgGaO$_4$. We pointed out that the
particle-hole excitation of a simple non-interacting
spinon Fermi sea already gives both the broad continuum and the upper 
excitation edge in the inelastic neutron scattering 
spectrum. In the future work, we will variationally 
optimize the energy against the trial ground state wavefunction 
that is constructed from a more generic spinon Fermi surface mean-field state. 
It will be ideal to directly compute the correlation function 
of the local moments with respect to the variational ground state.

To summarize, we have provided a strong evidence of
the anisotropic spin interaction and strong SOC 
in YbMgGaO$_4$. In particular, the nearest-neighbor spin model,
as used throughout the paper, proves to be the appropriate 
description of the system.

\emph{Acknowledgements.}---G.C.  sincerely acknowledges Dr. Martin Mourigal for discussion
and appologizes for not informing him properly about adapting Fig.~\ref{fig2}a. 
We thank Dr. Zhong Wang at IAS of Tsinghua University 
and Dr. Nanlin Wang at ICQM of Peking University 
for the hospitality during our visit in August 2016. 
This work is supported by the Start-up Funds of Fudan University 
(Shanghai, People's Republic of China) and the Thousand-Youth-Talent 
Program (G.C.) of People's Republic of China.

\bibliography{refs_Aug}

\end{document}